\documentstyle[a4,epsfig,11pt]{article}
\begin{document}

\baselineskip=7mm
\def\ap#1#2#3{           {\it Ann. Phys. (NY) }{\bf #1} (19#2) #3}
\def\arnps#1#2#3{        {\it Ann. Rev. Nucl. Part. Sci. }{\bf #1} (19#2) #3}
\def\cnpp#1#2#3{        {\it Comm. Nucl. Part. Phys. }{\bf #1} (19#2) #3}
\def\apj#1#2#3{          {\it Astrophys. J. }{\bf #1} (19#2) #3}
\def\asr#1#2#3{          {\it Astrophys. Space Rev. }{\bf #1} (19#2) #3}
\def\ass#1#2#3{          {\it Astrophys. Space Sci. }{\bf #1} (19#2) #3}

\def\apjl#1#2#3{         {\it Astrophys. J. Lett. }{\bf #1} (19#2) #3}
\def\ass#1#2#3{          {\it Astrophys. Space Sci. }{\bf #1} (19#2) #3}
\def\jel#1#2#3{         {\it Journal Europhys. Lett. }{\bf #1} (19#2) #3}

\def\ib#1#2#3{           {\it ibid. }{\bf #1} (19#2) #3}
\def\nat#1#2#3{          {\it Nature }{\bf #1} (19#2) #3}
\def\nps#1#2#3{          {\it Nucl. Phys. B (Proc. Suppl.) } {\bf #1} (19#2) #3}
\def\np#1#2#3{           {\it Nucl. Phys. }{\bf #1} (19#2) #3}

\def\pl#1#2#3{           {\it Phys. Lett. }{\bf #1} (19#2) #3}
\def\pr#1#2#3{           {\it Phys. Rev. }{\bf #1} (19#2) #3}
\def\prep#1#2#3{         {\it Phys. Rep. }{\bf #1} (19#2) #3}
\def\prl#1#2#3{          {\it Phys. Rev. Lett. }{\bf #1} (19#2) #3}
\def\pw#1#2#3{          {\it Particle World }{\bf #1} (19#2) #3}
\def\ptp#1#2#3{          {\it Prog. Theor. Phys. }{\bf #1} (19#2) #3}
\def\jppnp#1#2#3{         {\it J. Prog. Part. Nucl. Phys. }{\bf #1} (19#2) #3}

\def\rpp#1#2#3{         {\it Rep. on Prog. in Phys. }{\bf #1} (19#2) #3}
\def\ptps#1#2#3{         {\it Prog. Theor. Phys. Suppl. }{\bf #1} (19#2) #3}
\def\rmp#1#2#3{          {\it Rev. Mod. Phys. }{\bf #1} (19#2) #3}
\def\zp#1#2#3{           {\it Zeit. fur Physik }{\bf #1} (19#2) #3}
\def\fp#1#2#3{           {\it Fortschr. Phys. }{\bf #1} (19#2) #3}
\def\Zp#1#2#3{           {\it Z. Physik }{\bf #1} (19#2) #3}
\def\Sci#1#2#3{          {\it Science }{\bf #1} (19#2) #3}

\def\n.c.#1#2#3{         {\it Nuovo Cim. }{\bf #1} (19#2) #3}
\def\r.n.c.#1#2#3{       {\it Riv. del Nuovo Cim. }{\bf #1} (19#2) #3}
\def\sjnp#1#2#3{         {\it Sov. J. Nucl. Phys. }{\bf #1} (19#2) #3}
\def\yf#1#2#3{           {\it Yad. Fiz. }{\bf #1} (19#2) #3}
\def\zetf#1#2#3{         {\it Z. Eksp. Teor. Fiz. }{\bf #1} (19#2) #3}
\def\zetfpr#1#2#3{         {\it Z. Eksp. Teor. Fiz. Pisma. Red. }{\bf #1} (19#2) #3}
\def\jetp#1#2#3{         {\it JETP }{\bf #1} (19#2) #3}
\def\mpl#1#2#3{          {\it Mod. Phys. Lett. }{\bf #1} (19#2) #3}
\def\ufn#1#2#3{          {\it Usp. Fiz. Naut. }{\bf #1} (19#2) #3}
\def\sp#1#2#3{           {\it Sov. Phys.-Usp.}{\bf #1} (19#2) #3}
\def\ppnp#1#2#3{           {\it Prog. Part. Nucl. Phys. }{\bf #1} (19#2) #3}
\def\cnpp#1#2#3{           {\it Comm. Nucl. Part. Phys. }{\bf #1} (19#2) #3}
\def\ijmp#1#2#3{           {\it Int. J. Mod. Phys. }{\bf #1} (19#2) #3}
\def\ic#1#2#3{           {\it Investigaci\'on y Ciencia }{\bf #1} (19#2) #3}
\def\tp{these proceedings}
\def\pc{private communication}
\def\ip{in preparation}
\newcommand{\TeV}{\,{\rm TeV}}
\newcommand{\GeV}{\,{\rm GeV}}
\newcommand{\MeV}{\,{\rm MeV}}
\newcommand{\keV}{\,{\rm keV}}
\newcommand{\eV}{\,{\rm eV}}
\newcommand{\Tr}{{\rm Tr}\!}
\renewcommand{\arraystretch}{1.2}
\newcommand{\be}{\begin{equation}}
\newcommand{\ee}{\end{equation}}
\newcommand{\bea}{\begin{eqnarray}}
\newcommand{\eea}{\end{eqnarray}}
\newcommand{\ba}{\begin{array}}
\newcommand{\ea}{\end{array}}
\newcommand{\bmat}{\left(\ba}
\newcommand{\emat}{\ea\right)}
\newcommand{\refs}[1]{(\ref{#1})}
\newcommand{\ler}{\stackrel{\scriptstyle <}{\scriptstyle\sim}}
\newcommand{\ger}{\stackrel{\scriptstyle >}{\scriptstyle\sim}}
\newcommand{\lag}{\langle}
\newcommand{\rag}{\rangle}
\newcommand{\ns}{\normalsize}
\newcommand{\cm}{{\cal M}}
\newcommand{\gr}{m_{3/2}}
\newcommand{\p}{\partial}
\renewcommand{\le}{\left(}
\newcommand{\ri}{\right)}
\relax
\def\321{$SU(3)\times SU(2)\times U(1)$}
\def\ord{{\cal O}}
\def\tl{{\tilde{l}}}
\def\tL{{\tilde{L}}}
\def\bd{{\overline{d}}}
\def\tL{{\tilde{L}}}
\def\a{\alpha}
\def\b{\beta}
\def\g{\gamma}
\def\c{\chi}
\def\d{\delta}
\def\D{\Delta}
\def\db{{\overline{\delta}}}
\def\Db{{\overline{\Delta}}}
\def\e{\epsilon}
\def\l{\lambda}
\def\n{\nu}
\def\m{\mu}
\def\nt{{\tilde{\nu}}}
\def\p{\phi}
\def\P{\Phi}
\def\sol{\Delta_{\odot}}
\def\atm{\Delta_{atm}}
\def\k{\kappa}
\def\x{\xi}
\def\r{\rho}
\def\s{\sigma}
\def\t{\tau}
\def\th{\theta}
\def\om{\omega}
\def\ne{\nu_e}
\def\nm{\nu_{\mu}}
\def\snui{\tilde{\nu_i}}
\def\ehat{\hat{\e}}
\def\la{{\makebox{\tiny{\bf loop}}}}
\def\ta{\tilde{a}}
\def\tb{\tilde{b}}
\def\mb{m_{1b}}
\def\mt{m_{1 \tau}}
\def\rl{{\rho}_l}
\def\meg{\m \rightarrow e \g}

\renewcommand{\Huge}{\Large}
\renewcommand{\LARGE}{\Large}
\renewcommand{\Large}{\large}
\title{
\hfill hep-ph/0203182\\[.2cm]
\hfill TIFR-TH/02-01\\
Bi-maximal Mixing and Bilinear R Violation}

\author{ Anjan S. Joshipura~$^{a}$, Rishikesh D. Vaidya~$^{a}$
and Sudhir K. Vempati\footnote{presently at INFN, Sezione di Padova,
via F. Marzolo 8, 35131 Padova, Italy.} ~$^{b}$ \\
{\ns\it $(a)$~ Theoretical Physics Group, Physical Research Laboratory,}\\
{\ns\it Navarangpura, Ahmedabad, 380 009, India.}\\
{\ns\it $(b)$~ Department of Theoretical Physics, Tata Institute of Fundamental
Research,}\\
{\ns\it Colaba, Mumbai 400 005, India. }}

\maketitle


\begin{abstract}
We perform a general analytic study of feasibility of obtaining a
combined explanation for the deficits in the solar and the
atmospheric neutrino fluxes with  two large mixing angles in
supersymmetric model with bilinear $R$ parity violations. The
required hierarchy among the solar and atmospheric neutrino mass
scales follows in this framework in the presence of an approximate
Higgs - slepton universality at the weak scale. The solar
mixing angle is shown to be related to  non-universality in
slepton mass terms specifically to differences in soft parameters
of the first two leptonic generations. It is shown that this
flavour universality violation should be as strong as the
Higgs-slepton universality violation if solar neutrino mixing
angle is to be large. The standard supergravity models with
universal boundary conditions at a high scale lead to the required
Higgs-slepton universality violations but the predicted violation
of  flavour universality  among the first two generations is much
smaller than required. This model therefore cannot provide an
explanation of large solar neutrino mixing angle unless some universality
violations in soft supersymmetry breaking parameters are
introduced at a high scale itself.

\end{abstract}

\section{Introduction}
Experimental observations of deficits in the solar \cite{sksolar} and atmospheric
\cite{skatm} neutrino fluxes have provided concrete ground to believe in
neutrino oscillations. These experimental results are consistent
with a simple picture of three active neutrinos mixing with each
other. Within this picture, two independent (mass)$^2$ differences
($\sol,\atm$) among three neutrinos govern the oscillations of the
solar and atmospheric neutrinos respectively. 
One needs $\sol/ \atm~\leq 10^{-2}$.
Two of the mixing angles determining
amplitudes of these oscillations are required to be large \cite{datanal}. The
third mixing angle measured by the survival probability of the
electron neutrinos in laboratory experiments such as CHOOZ is
found to be much smaller $\leq 0.1$ \cite{choose}.

Different theoretical possibilities have been suggested for
obtaining the above neutrino spectrum with two large mixing
angles \cite{revs}. One potentially interesting possibility in this
regard is supersymmetric standard model containing bilinear $R$ parity and
lepton number violation
\cite{hall,hempfling,asjbabu,asjskv,bilinear_univ,bilinear_nouniv,kaplan,romao,yama}.
The following features of the model make
it an ideal candidate for the description of neutrino masses. (1)
The lepton number violations and hence neutrino masses and mixing
are described in this model in terms of three parameters. Ratios
of these parameters control neutrino mixing which can be naturally
large. (2) The mechanism for suppression of neutrino masses
compared to other fermion masses is automatically built-in for two of
the most popular supersymmetry breaking scenario namely the
minimal supergravity model (mSUGRA) and models with gauge mediated
supersymmetry breaking (GMSB). Extensive studies of these models
have been carried out in the literature \cite{hempfling,asjbabu,asjskv,bilinear_univ}.
Our aim in this paper is
to discuss under what conditions the bilinear model can lead to
two large mixing angle among neutrinos. We discuss this issue
analytically and in the process show that the two scenarios
mentioned above cannot lead to two large mixing angles although
small angle mixing solution to the solar neutrino problem is
possible\footnote{Feasibility of only small mixing angle solution
was pointed out also in \cite{yama}. Our analysis considerably differs
from theirs.}.

The suppression in neutrino masses in mSUGRA and GMSB arises due to
equality at a high scale($\equiv M_X$) of soft parameters of one
of the Higgs fields ($\equiv H_1'$) with the corresponding
parameters of the leptonic doublets having the same quantum
numbers as $H_1'$. Small differences arise in these equal
parameters at the weak scale due to RG scaling. For example,
one finds in case of mSUGRA
\be \label{deltam} \Delta m_i^2\equiv
(m_{\snui'}^2-m_{H'_1}^2)\approx {3 h_b^2\over 4 \pi^2}
\ln{M_X\over M_Z}m_{susy}^2\approx 2 \cdot 10^{-3}~ m_{susy}^2  \ee
where $m_{\nu_i'}^2(i=1,2,3),m_{H_1'}^2$ respectively denote the
weak scale values of the soft SUSY breaking masses of the
sneutrino and $H_1'$ respectively and $m_{SUSY}$ is the typical
SUSY breaking scale $\sim \mathcal{O}$(100 GeV) . The $h_b$ in the
above equation
refers to the $b$-quark Yukawa coupling. The neutrino masses in
this model involve the above and similar differences among $B$
parameters. The suppression in these differences leads to
suppression in neutrino masses. Thus the smallness of neutrino
masses is linked to near universality of the {\it Higgs} ($H_1'$)
and {\it sneutrino} soft parameters. As we will discuss in this
paper, the solar neutrino mixing angle is directly linked to { \it
flavour} universality violation, \textit{i.e,} to differences in
sneutrino mass parameters themselves. More specifically, the solar
neutrino mixing angle involves the  parameter
\be \label{flavourdiff}
\delta=\frac{m_{\nu_2'}^2-m_{\nu_1'}^2}{\Delta m_1^2+\Delta m_2^2} ,
\ee
which is required to be  $\ord(1)$ implying that the weak scale
universality violation among different flavours are required to be
as strong as the corresponding  Higgs-slepton universality
violations. This is in sharp contrast with the expectations based
on mSUGRA and GMSB where the former violations are mainly
controlled by the muon Yukawa coupling while the latter by the $b$
or $\tau$ Yukawa couplings. Thus $\d$ in eq.(\ref{flavourdiff}) is
of $\ord(10^{-4})$ instead of being one.

Link between universality violation and large mixing was brought
out in the numerical study of \cite{romao}. In contrast to their work,
our analytical study allows us to determine specific pattern of
universality violation and also allows us to quantify the amount
of violation needed to obtain the LMA solution for the solar
neutrino problem.

We present our results in the following manner. The next section
outlines general formalism we adopt and our assumptions. It also
contains analytic  discussion of neutrino mixing and masses in
this scheme. The close link between large angle solar neutrino
solution and flavour violation is emphasised in section (3) which
also contains results based on numerical analysis. The last
section contains a summary. Some of the technical aspects relevant
to discussions in the text are elaborated in the appendices.
\section{Sources of neutrino masses}
In this section, we derive analytical conditions on the low energy
susy parameters in order to have a phenomenologically consistent 
neutrino spectrum. To this extent we do not assume any specific
structure of the soft masses. Further assumptions regarding various
contributions to neutrino mass spectrum within mSUGRA inspired
scenarios are discussed as and where required. 
The superpotential takes the following form in our case:
 \be \label{sup}
 W = h^u_{ij} Q_i u_j^c H_2
+ h^d_{i} Q_i d_i^c H_1' + h^e_{i} L_i' e_i^c H_1'  + \m' H_1'H_2
+ \e_i L_i' H_2 . \ee
Without loss of generality, we have chosen above the basis in
which the down quarks and charged lepton masses are diagonal. The
$\e_i$ characterise lepton number violation in this basis.

We have the following soft supersymmetry breaking terms at the weak scale: 
 \bea \label{soft}
V_{soft}& = & m_{H_1^{\prime 0}}^2 |H_1^{' 0} |^2 + m_{H_2^{0}}^2
|H_2^{0}|^2 + m_{\snui'}^2 |\snui'|^2 \nonumber \\
&-& \left( B_\m \m' H_1^{'0} H_2^0 + c. c \right) -  B_i \e_i
\left( \snui' H_2^0 + c. c. \right) + \ldots \eea

\noindent
Note that the above equation refers to soft terms at the weak
scale. For simplicity we have displayed only the terms involving
neutral fields in the above equation. The following comments are
needed in connection with eq.(\ref{soft}):\\
\noindent ({\it i}) Although we have allowed for arbitrary
diagonal sneutrino masses, we have not included off-diagonal
sneutrino masses in this primed basis since such off-diagonal
masses are severely constrained by flavour violating processes,
e.g.  $\mu\rightarrow e\gamma$ \cite{gabbiani}. \\
\noindent ({\it ii}) $V_{soft}$ does not contain sneutrino-Higgs
mixing terms of the form $ m^2_{\tilde{\nu'}_i H'_1}\snui'^* H_1'$
although they are allowed by the gauge symmetry. Such terms are
not present in the minimal supergravity theory at  high scale. The
renormalization group (RG) equations for $m^2_{\snui' H_1'}$ given
in the appendix, eq.(\ref{mnuh1rge}) show that these terms cannot
get generated even at the weak scale if they are not present at
high scale. Thus it is meaningful to omit these terms. \textit{We
should emphasise that this statement is very specific to the
particular basis in which bilinear terms are not rotated away from
the superpotential until the weak scale and neglect of such terms
would not be justified in any other basis.} In our case, the
$\snui^* H_1$ term would make its appearance when we go to the
basis with no bilinear $R$ violating terms in the superpotential
at the weak  scale.

The neutrino masses arise from several sources in this model.
Discussion of these sources becomes transparent if we re-express
eq.(\ref{sup}) in the new basis in which bilinear terms are
rotated away from $W$\footnote{Note that this definition of a new
basis is same as that of Ref.\cite{carlos}. However in the present
work, this rotation is done  only at the weak scale in contrast
to \cite{carlos} where it is scale-dependent.}:

\bea
H_1 &=&  { \m' H_1' + \sum_i \e_i L'_i \over \m'}, \nonumber \\
\label{basis1} L_i &=&  { \m' L'_i - \e_i H'_1 \over \m'}. \eea
\noindent This basis are simple but are orthonormal only up to
$\ord(\frac{\e^2}{\m'^2})$. This approximation is sufficient for
most of our discussions since $\e_i$ are required to be much
smaller than the typical SUSY scale $\m'$ in order to reproduce
the scale of neutrino masses correctly. Generalisation of
eq.(\ref{basis1}) valid to higher  order in $\e_i$ and its
consequences are discussed in the appendix B. Eq.(\ref{sup}) takes
the following form in the unprimed basis:
\be\label{suprotated}
W = h^u_{ij} Q_i u_j^c H_2 + h^d_{i} Q_i d_i^c H_1 + h^e_{i} L_i
e_i^c H_1 - \l'_{ijk} L_i Q_j d_k^c - \l_{ijk} L_i L_j e_k^c + \m
H_1 H_2 .\ee
where
\bea
\m^2&=&\m'^2+\e_1^2+\e_2^2+\e_3^2\approx \m'^2~,\nonumber \\
\l'_{ijk} &=& {\e_i \over \m'}~ h^d_j\d_{jk} ~, \nonumber \\
\label{lambda}
\l_{ijk}&=&(\d_{ik}h^e_i\frac{\e_j}{\m'}-\d_{jk}h^e_j\frac{\e_i}{\m'})
~. \eea
Similarly, after rotating primed terms in eq.(\ref{soft}) and
adding the contribution of the supersymmetric part, we get the following
expression  for the full scalar potential in the unprimed basis:
\bea
V_{scalar}& = & (m_{H_1^{\prime 0}}^2+\m^2)
|H_1^0 |^2 + (m_{H_2^0}^2+\m^2) |H_2^0|^2 + m_{\snui'}^2 |\snui|^2
+ \Delta m_i^2 {\e_i \over \m} \left( \snui^{\star} H_1^0 + c.c
\right) \nonumber \\
\label{scalar}
&-& \left( B_\m \m H_1^0 H_2^0 + c. c \right)
- \Delta B_i \e_i \left( \snui H_2^0 + c. c \right)\nonumber \\
& + & {1 \over 8}
(g^2 + g'^2) \left( |H_1^0|^2+|\snui|^2 - |H_2^0|^2 \right)^2 ~,
\eea
where
\be \ba{cc}\Delta m_i^2\equiv  m_{\snui'}^2-m_{H_1^{\prime
0}}^2~~~&~~~ ~~ \Delta B_i \equiv B_i-B_\m ~.\\ \ea \ee

Two major sources of neutrino masses arise from
eqs.(\ref{lambda},\ref{scalar}). Minimization of eq.(\ref{scalar})
generates sneutrino \textit{vev}:
\be \label{omega}<\nu_i>\equiv \e_i k_i \ee
where \be \label{ki} k_i\approx \frac{v_1}{\m}\frac{(-\Delta
m_i^2+\tan\b~ \m\Delta B_i)}{(m_{\snui}^2+1/2 M_Z^2 \cos~2\b)} \ee
$v_1=<H_1^0>$ and  $M_Z$ represents the $Z$ boson mass. Sneutrino
\textit{vevs} lead  to neutrino masses through their mixing with
neutral-gauginos:
\be \label{treem} {\cal M}_{\mathbf{tree}}\equiv
A_0<\tilde{\nu_i}><\tilde{\nu_j}>=A_0\e_i\e_jk_ik_j~.\ee $A_0$ is
obtained by diagonalizing the $7 \times 7$ neutrino-neutralino mass
matrix in the standard way \cite{asjmarek}:
 \be \label{m0}
A_0 = { \m \le g^2 + g'^2 \ri \over 2 \le -c \m M_2 +  M_W^2
\sin 2 \b \le c + \tan^2 \th_W \ri \ri} ~, \ee
where $\th_W$ represents the Weinberg angle and $M_W$ represents the
W-boson mass. $c$ is given by $5 g^2/ 3 g^{'~2} \sim 0.5 $ with $M_2$
representing the standard gaugino mass parameter.

The trilinear terms in eq.(\ref{lambda}) lead  to the second
contribution to neutrino masses at 1-loop level. Since these
couplings are proportional to the Yukawa couplings, the dominant
contributions arise due to exchanges of the $b$-quark-squark and
$\tau$-lepton-slepton in the loops. The loop induced mass matrix
is of the form :
\be \label{loop}
 ({\cal M}_{loop})_{ij} = \e_i\e_j\le A_b+ A_\tau   \le 1 -
\delta_{i3}
\ri \le 1 - \delta_{j3} \ri \ri
 \ee
where
\bea \label{abtau}
A_b &=& {3 \over 16 \pi^2 } {v_1 \over \m^2} h_b^3 \sin \phi_b
\cos \phi_b ln \le M_{2b}^2 \over M_{1b}^2 \ri, \\
A_{\tau} &=& {1 \over 16 \pi^2 } {v_1 \over \m^2} h_\tau^3 \sin
\phi_\tau \cos \phi_\tau ln \le M_{2\tau}^2 \over M_{1\tau}^2
\ri.\eea

\noindent
Here  $\phi_{b,(\tau)}$ denotes mixing between the left and the
right handed squark (sneutrino) fields. These mixing angles are
proportional to the $b$ and $\tau$ Yukawa couplings. Approximating
them by $\frac{m_{b,\tau}}{m_{\mathbf{susy}}}$ we get the
following numerical values
\bea
A_0&\approx& 5\cdot 10^{-3} \GeV^{-1}\nonumber, \\
A_b&\approx& 3\cdot 10^{-10} \GeV^{-1}\nonumber, \\
\label{numeric}
A_{\tau}&\approx&4 \cdot 10^{-12}\GeV^{-1}. \eea
for $m_{susy}\sim 100\GeV.$ There are other loop
contributions to neutrino masses and a complete discussion is given
in \cite{hempfling,romao,chun1loop}. We have retained here only those
contributions which are known \cite{chun1loop} to be dominant in
case of mSUGRA and GMSB. The additional contributions not included
in the text come from, (a) $R$ parity violating mixing of the
charged sleptons with Higgs fields (b) sneutrino exchange diagrams 
through R-parity violating sneutrino-Higgs mixing and 
(c) loop contribution to the tree level neutrino
neutralino mixing. These contributions are sub-dominant as long as
the parameters $\Delta m_i^2,\Delta B_i$ are suppressed
\cite{chun1loop}. Such suppression is required purely from the
phenomenological point as we argue below. It is then consistent to
omit these sub-dominant terms for the analytical discussion that
follows. We however discuss these additional contributions in the
appendix B.

The total neutrino mass matrix is given by
\be \label{mtot}
 \le{\cal M}_{tot}\ri_{ij}=A_0 \e_i\e_j k_i k_j+
\e_i\e_j\le A_b+ A_\tau \le 1 - \delta_{i3} \ri \le 1 -
\delta_{j3} \ri \ri ~. \ee
The desired hierarchy among neutrino masses is automatically built
in the above equations in view of typical numerical values of the
parameters $A_{0,b,\tau}$. The tree contribution dominates over
the rest (unless $k_i$ are enormously suppressed) but it leads to
only one massive neutrino. Switching on the b-quark contribution
gives mass to the other neutrino, one neutrino still remaining
massless at this stage. The latter obtains its mass from somewhat
less dominant contribution due to $A_{\tau}$. Note that hierarchy
among the first two neutrino masses need not be very strong due to
similar magnitudes of $A_{b,\tau}$. The above statements are made
explicit below which also contains discussion on neutrino mixing.
\subsection{Neutrino masses and mixing}
The tree-level neutrino mass matrix can be easily diagonalized:
\be \label{treed}
U_0~{\cal M}_{tree}U_0^T= diag \{0, 0, m_{\nu_3} \}~, \ee
where
\be \label{treemix} U_0^T = \bmat{ccc} c_{2}& s_{2} c_{3} &s_{2}
s_{3} \\ -s_{2}& c_{2} c_{3} & c_{2} s_{3} \\ 0& - s_{3} & c_{3}
\\ \emat ~,\ee
with $s_{2,3}=\sin\theta_{2,3}$ and
\be \ba{cc} \tan\theta_{2}= \e_1 \k_1 / \e_2k_2 ~;&
\tan\theta_3=\sqrt{\e_1^2k_1^2+\e_2^2k_2^2}/ \e_3k_3\\
\ea~.\ee

\noindent
The total mass matrix eq.(\ref{mtot}), assumes the following form
in basis with diagonal tree mass matrix:

\be \label{rotm}
 U_0~{\cal M}_{\mathbf{tot}}~U_0^T=\bmat{ccc}
a_1^2(A_b+A_{\tau}) &a_1 (A_b a_2+A_\tau b_2)&a_1 (A_b a_3+A_\tau
b_3)\\
a_1(A_b a_2+A_\tau b_2)& A_b a_2^2+A_\tau b_2^2&A_b a_2 a_3+A_\tau b_2 b_3\\
a_1(A_b a_3+A_\tau b_3)&A_b a_2 a_3+A_\tau b_2 b_3&A_0\omega^2+A_b a_3^2+A_\tau b_3^2\\
 \emat, \ee
where \bea
a_1&=&{\e_1\e_2\over \omega_{\perp}}(k_1-k_2), \nonumber \\
a_2&=&{\e_3\over \omega_{\perp}\omega}(\e_1^2 k_1(k_1-k_3)+ \e_2^2
k_2 (k_2-k_3)), \nonumber \\
a_3&=&-{1\over \omega}(\e_1^2 k_1+\e_2^2 k_2+\e_3^2 k_3), \nonumber \\
b_2&=&{\e_3 k_3\over \omega_{\perp}}~ b_3, \nonumber \\
\label{abs}
b_3&=&-{1\over \omega} (\e_1^2
k_1+\e_2^2 k_2)~, \eea with
\bea \omega&=&(\e_1^2 k_1^2+\e_2^2 k_2^2+\e_3^2 k_3^2)^{1/2}
\nonumber, \\
\omega_{\perp}&=& (\e_1^2 k_1^2+\e_2^2 k_2^2)^{1/2} ~.\eea
The subsequent diagonalization can be approximately done if we
neglect terms of $\ord(\frac{A_{b,\tau}}{A_0})$. Let
\be \label{s1} U_1^T = \bmat{ccc}
c_1 & s_1 & 0  \\
-s_1 & c_1 & 0  \\
0 & 0 & 1 \emat ~,\ee
where \be \label{theta1} \tan 2 \th_1 = {2 a_1 \le A_b a_2 +
A_\tau b_2 \ri \over A_b \le a_2^2 - a_1^2 \ri + A_\tau \le
b_2^2 - a_1^2 \ri }. \ee
We then have
\be U_1U_0{\cal M}_{tot}U_0^T U_1^T = \bmat{ccc}
m_{\nu_1} & 0 & 0\\
0 & m_{\nu_2} & 0\\
0 & 0& m_{\nu_3} \emat +\ord(\frac{A_{b,\tau}}{A_0}). \ee
\noindent
The eigenvalues are approximately given by
\bea
m_{\nu_1}&\approx& A_\tau ~ {a_1^2 (a_2 - b_2)^2 \over  (a_1^2+a_2^2)},\nonumber \\
m_{\nu_2}&\approx& A_b~ (a_1^2+a_2^2),\nonumber \\
\label{numasses}
m_{\nu_3}&\approx& A_0~ \omega^2. \eea

\noindent
The mixing among neutrinos is described by
\be \label{u} U\equiv U_0^T~U_1^T=
\bmat{ccc} c_2c_1-s_1s_2c_3&c_2 s_1+c_1s_2c_3&s_2s_3\\
-s_2c_1-s_1c_2c_3&-s_2 s_1+c_1c_2c_3&c_2s_3\\
s_1s_3&-c_1s_3&c_3\\ \emat ~.\ee

\noindent
Let us now discuss consequences of the above algebraic results.

\noindent (1) It follows from eqs.(\ref{numeric},\ref{numasses})
that the neutrino masses obey the desired hierarchy:
$$ m_{\nu_1}\ler m_{\nu_2}\ll m_{\nu_3}. $$
\noindent (2) The neutrino masses relevant for the solar and
atmospheric  scales are respectively given by $A_0 \e^2k^2$ and
$A_b \e^2$ leading to
$$\frac{\sol}{\atm}\approx
\left(\frac{A_b}{A_0}\right)^2\frac{1}{k^4}~,$$
where $\e, k$ represent typical values of $\e_i,k_i$. It follows
that the ratio of the solar to atmospheric scales is {\it
independent}  of the $R$ violating parameters $\e_i$ and depends
upon the values of the soft parameters represented by $k$. One
typically needs
\be \label{ek} \e\sim 10^{-1} \GeV~~~;~~~ k\sim 10^{-3}-10^{-4}  \ee
in order to reproduce the scales correctly. This shows in
particular that irrespective of details of the SUSY breaking the
Higgs-slepton universality (corresponding to very small values of
$k$ ) is unavoidable in this model if neutrino masses are to be
correctly reproduced.

\noindent (3) A particularly interesting limit would be the case
where $k \sim {\mathcal O}(1)$. This is what one expects in a 
truly non-universal regime of soft masses \cite{murayama}. From the
above, we see that in this limit, one can infact choose $\e_i \sim
\mathcal{O}$(MeV) and generate neutrino masses of $\mathcal{O}$(eV).   
However only one neutrino would be massive in this scenario, with the
the other neutrinos being with extremely negligible masses and without
any relevance for phenomenology. 

\noindent (4) If exact flavour universality were to hold between
the first two generations then $k_1=k_2$ (see eq.(\ref{ki})). In
this case $a_1$ as defined in eq.(\ref{abs}) would be zero leading
to $s_1=0$ in eq.(\ref{s1}) . The $s_1$ is required to be large in
order to obtain the large mixing angle solution and obtaining this
solution would need very sizable departures from the flavour
universality among the first two generations. We quantify these
remarks in the next section.

\section{Neutrino mixing and departure from flavour universality}

We derived approximate expressions for the neutrino masses and
mixing without any specific assumption on the soft symmetry
breaking sector. The entire neutrino spectrum can be parameterized
in terms of three $\e_i$ and three $k_i$ of which $k_i$ depend
upon the soft SUSY breaking parameters. We now quantify the amount
of flavour universality violations needed for obtaining the most
preferred large angle solution to the solar neutrino problem. The
following two parameters are introduced as a measure of
universality violation:
\be \label{xy}
\ba{cc}
x= (k_1-k_3)/(k_1+k_3) &;~~ y=(k_1-k_2)/(k_1+k_2) \\ \ea
\ee
We regard $x$ and $y$ as independent parameters but restrict their
variation to values between (-1,1) in the numerical analysis that
follows.

The neutrino mixing is determined by the matrix $U$ in
eq.(\ref{u}). Due to hierarchical mass spectrum, the survival
probabilities for the solar and atmospheric neutrinos
approximately assume two generation form. The  corresponding
mixing angles $\theta_{\odot}$ and $\theta_{\mathbf{atm}}$
are given in terms of elements of the mixing matrix $U$ as
follows:

\bea
\sin^2 2\theta_{atm}&\sim& 4~U_{\mu 3}^2(1-U_{\mu 3})^2\approx
0.8-1.0 \nonumber \\
\sin^2 2\theta_{\odot}&\sim&4~U_{e2}^2U_{e1}^2\approx 0.75-1.0 \nonumber \\
\label{thaths} \sin^2\theta_{CHOOZ}&\sim&U_{e3}^2 \leq 0.01~, \eea
where numbers on the RHS correspond to the required values for
these parameters based on two generation analysis of the
experimental data \cite{datanal}.

We can convert the above restrictions on $\theta_{\odot},
\theta_{\mathbf{atm}}$ to restrictions on the mixing angles
$s_{1,2,3}$ entering the definition of $U$. The CHOOZ result
requires $|s_2s_3|\leq 0.1$ and the nearly maximal atmospheric
mixing is obtained with $|c_2s_3|\approx \frac{1}{\sqrt{2}}$. This
requires small $s_2$ and large $s_3$. The solar neutrino mixing angle
defined in eq.(\ref{thaths}) coincides with $s_1$ in this limit.
We thus need $\sin^2 2\theta_1\sim 0.75-1.$ Large value of $s_1$
in turn needs sizable departure from flavour universality as
will be argued in the last subsection.

The expressions for mixing angles and masses obtained in the last
section can be used to approximately determine the allowed ranges
of parameters $k_i,\e_i$ which explain the solar and atmospheric
neutrino anomalies. We approximately need  $|s_2|\leq \sqrt{2}
U_{e3}$ and $|s_3| \approx \frac{1}{\sqrt{2}}$. This implies:
\bea
\e_2^2k_2^2\approx \e_3^3 k_3^2~, \nonumber \\
\label{e2k2} |\e_1k_1|\approx \sqrt{2}|U_{e3}\e_2k_2| .\eea
The magnitude of $\e_3k_3$ is then approximately fixed by the
atmospheric mass scale: \be \label{e3k3}
\e_3^2k_3^2\approx \frac{m_{\nu_3}}{2 A_0}\approx
{\sqrt{\atm}\over 2 A_0}~, \ee
while the solar scale and mixing angle determines $\e_3^2$:
\be \label{e3} \e_3^2\approx\left({\sqrt{\sol}\over 2 A_b
\cos^2\theta_{\odot}}\right) {(1+x)^2(1-y)^2\over (x-y)^2} ~. \ee

Eqs.(\ref{e2k2},\ref{e3k3},\ref{e3}) allow us to express
magnitudes of all $\e_i,k_i$ in terms of $x,y$ , approximately known
$A_{0,b}$ and the experimentally measurable quantities.

The solar neutrino mixing angle following from eq.(\ref{theta1}) is given
in the limit $A_{\tau}\ll A_b$ by
\be \label{solang} \tan^2\theta_{\odot}\approx
\tan^2\theta_1\approx {4 U_{e3}^2 y^2 (1-x)^2\over (x-y)^2} ~. \ee
We have used eq.(\ref{e2k2}) in deriving the above relation. It is
clear that large $\theta_1$ require sizable departure from flavour
universality, i.e. sizable $y$. Moreover, one typically needs
$|x-y|\approx 2 |U_{e3} y (1-x)|$ in order to obtain a sizable
solar neutrino mixing angle.

We now numerically determine the region in the $x,y$ plane needed
to reproduce the required ranges in mixing angle and masses. We
make use of  eqs.(\ref{e2k2}-\ref{e3}) to determine the
approximate input values of $\e_i,k_i$ in terms of the $\sol,\atm$
and $x,y$. We allow input values to vary by varying $\sol,\atm$
over the experimentally allowed ranges. We also randomly choose
$x,y$ between -1 and 1. Through this procedure, we choose a set of
$1.5 \times 10^5$ different values for the input parameters $\e_i,k_i$. Then we
numerically diagonalize the total neutrino mass matrix,
eq.(\ref{mtot}) for each of these values of $\e_i,k_i$ and
determine a set of $x,y$ values which correctly reproduces the
allowed ranges of the solar and atmospheric neutrino parameters
and lead to $|U_{e3}|\leq 0.1$. We obtain about 2024  $x,y$
values leading  to the correct description of neutrino anomalies.
These points in the $x,y$ plane are displayed in Fig.(1). This
figure, based on complete diagonalization clearly shows the
features obtained through approximate formulas. All the allowed
values of $x$ and $y$ are in the range $-0.9- -0.6$ and sizable
departure from universality is clearly seen. Also most points
satisfy approximate equality $|x-y|\sim 2 U_{e3} y$ needed to
obtain large solar neutrino mixing angle. As an illustration, we give below a
typical set of $\e_i,k_i$ which correctly reproduces all the
parameters:
\be \label{all}
\ba{cc} \e_3\sim 0.1 \GeV&;~ k_3\sim 1.1 ~\cdot 10^{-3} \\
\e_2\sim 0.031 \GeV&;~ k_2\sim 3.5 ~\cdot 10^{-3}\\
\e_1\sim 0.087\GeV&;~ k_1\sim 9.1 ~\cdot 10^{-4} \\ \ea \ee
Typically, one needs $\e_i\sim \ord(10^{-1}~GeV)$ and $k_i\sim
10^{-3}$ as argued before.

Let us now compare above phenomenological restrictions with
expectations based on specific framework like mSUGRA. In order to
obtain correct neutrino masses one needs parameters $k_i$
(\ref{ki}) to be suppressed, typically $k\sim 10^{-3}-10^{-4}$ as in
eq.(\ref{ek}). The other constraint is that $y$ should be
$\ord$(1). The $k_i$ provide a measure of the Higgs-slepton
universality violation. Typical value of $k_i$ obtained in mSUGRA
follows from eq.(1) and is in the range required from
phenomenology. Thus mSUGRA provides a very good framework to
understand neutrino mass hierarchy as has been demonstrated in
number of papers through detailed numerical calculations
\cite{hempfling,bilinear_univ,romao}. However mSUGRA would not be
able to provide the required value of $y$. This can be seen as
follows. Theoretically, $y$ can be approximately written using
eq.(\ref{ki}) as follows: \be \label{xvalue}
y\approx {\m \tan \b
(B_1-B_2)-(m_{\tilde{\nu'}_1}^2-m_{\tilde{\nu'}_2}^2)\over
\m\tan\b (\Delta B_1+\Delta B_2)-(\Delta m^2_1+\Delta m^2_2)}~,\ee
where we have neglected terms of order $(\Delta m_i^2)^2,(\Delta
B_i)^2$ etc. Within mSUGRA, $y$ is identically zero at the high
scale as $B_1=B_2$ and $m_{\nu_1}^2 = m_{\nu_2}^2$ due to the
universal boundary conditions. At the weak scale, this
universality condition is broken solely by RG evolution. In the
limit of neglecting first two generation Yukawa couplings, $y$ is
identically zero even at the weak scale. A rough estimate of
parameters appearing in $y$ can be obtained by approximately
integrating the RG equations, eqs.(\ref{dmlrge},\ref{dblrge})
given in appendix A. We see that

\bea
{m_{\tilde{\nu'}_1}^2-m_{\tilde{\nu'}_2}^2\over \Delta m_1^2+\Delta m_2^2}\approx
\frac{1}{6} \left( {m_\m\over m_b} \right)^2 \approx 10^{-4}~,\nonumber \\
\label{rgvalues} {(B_1-B_2)\over \Delta B_1+\Delta B_2}\approx
\frac{1}{6} \left({m_\m \over m_b} \right)^2 \approx 10^{-4}~.\eea
Together they would imply very small value for $y~ \sim ~0 $
instead of the required value of  $\ord$(1). Thus universal
boundary conditions of mSUGRA cannot lead to a large mixing angle
solution to the solar neutrino problem.

One need not consider a generic non-universal scenario to introduce 
universality violations. It is clear from the forgoing discussion that 
one \textit{only} needs small Higgs-slepton universality violation as well as 
flavour violation of similar magnitude to obtain a large solar neutrino
mixing angle. These violations can come from either non-universal slepton
mass terms or from non-universal B-terms or both. It is possible that 
such a scenario can arise from a higher theory of flavour either based
on string theory \cite{andrea} or through abelian \cite{dvali} or 
non-abelian \cite{anna} flavour symmetries. However, for a phenomenological
understanding it is clear that the existence of Higgs-slepton and flavour
universality violations at the high scale would lead to correct neutrino
mass spectrum at the weak scale. 
Knowing the value of $x$ and $y$ required for a correct neutrino
spectrum at the weak scale, it is possible to estimate the amount
of non-universality required at the high scale. For example, using
$y$, we have in the limit of neglecting contributions from $\Delta
B$ terms, the required slepton flavour universality violations to
be of order: \be m_{\tilde{\nu}'_2}^2(0) - m_{\tilde{\nu}'_1}^2(0)
\approx y ( m_{\tilde{\nu'}_2}^2(0) +  m_{\tilde{\nu'}_1}^2(0) ) +
2~y~ { m_{\tilde{\nu'}_2}^2(0)~ m_{\tilde{\nu'}_1}^2(0) \over
m_{H_1^0}^2(0) + \delta m_{H_1^0}^2 } \ee where $\delta
m^2_{H_1^0}$ represents the correction to the high scale  Higgs
mass due to RG scaling. From the above we see that for a large
negative $y$, $m_{\tilde{\nu}_1}^2(0)$ should be at least a factor
of 3 times larger than $m_{\tilde{\nu}_2}^2(0)$. Introducing such a
non-universality at the high scale would lead to the correct neutrino
spectrum at the weak scale. 

A similar analysis can also be considered for the B-terms \cite{asjskv}.
Moreover, such a pattern can be incorporated naturally in non-minimal
models of GMSB \cite{dine}. In these models, identical gauge quantum 
numbers of all sneutrinos assure almost universal sneutrino masses at the 
weak scale as in the case of mSUGRA. In contrast, there is no natural reason 
within these models for the flavour universal $B$ parameters. In fact, 
the $B$ parameters are assumed to vanish  in the minimal version of the scheme
\cite{mmm,borzu}. Thus the universality of B parameters at supersymmetry 
breaking scale holds by default. It is possible to choose non-universal 
and non-zero  $B_{1,2}$ terms to start with in this model. This does not 
significantly influence the conventional phenomenology of the minimal version 
as long as the parameters $\e_i$ are  much smaller than the $\mu$-parameter 
in the superpotential. But it allows the LMA solution as has been
demonstrated through a detailed numerical work \cite{asjskv}.

From the above discussion we see that the phenomenological requirement
of $k \sim {\mathcal O}(10^{-3})$ and $x,y \sim \mathcal{O}$(1) leads
to a specific class of non-universality at the high scale. A generic
non-universal soft spectrum might lead to much larger class of solutions.
Such an analysis is beyond the scope of present work.

\section{Comments}
Supersymmetric model with bilinear $R$ parity violations provides
a potentially interesting framework to study neutrino masses and
mixing. The dominant sources of neutrino masses can be
parameterized in this scenario in terms of three dimensionful
parameters $\e_i$ and three dimensional parameters $k_i$. The
$k_i$ depend on the structure of soft supersymmetry breaking terms
at the weak scale. We have tried to obtain phenomenological
restrictions on $\e_i$ and $k_i$ without making specific
assumptions on the values of the soft supersymmetry breaking
parameters. While neutrino masses can be suppressed by lowering
the overall scale $\e_i$ of $R$ parity violation,
phenomenologically preferred hierarchy in neutrino masses require
that both $\e_i$ and $k_i$ are suppressed, see eq.(\ref{ek}).
$k_i$ provide a measure of the Higgs-slepton universality and
suppression in their values indicate very small amount of this
violation. Such violation of universality is already built in the
mSUGRA and GMSB scenario.

A large solar neutrino mixing angle can be obtained consistently
within these scenarios only if flavour universality violations in
the soft parameters of the first two generations are almost as
large as the violation of Higgs-slepton universality. This feature
does not emerge in models where these universality violations are
generated solely by RG scaling as in the case of mSUGRA. Thus
mSUGRA seems more suitable to describe the less preferred small
mixing angle solution to the solar neutrino problem.

We concentrated throughout on the most dominant sources of
neutrino masses in this theory. This is a good assumption in case
of small universality violation. The other sources of neutrino
masses would become important in case of large universality
violation. It is not unlikely that these contributions could also
lead to a large solar neutrino mixing angle in such scenarios.

Another way to achieve the Large Mixing Angle solution is to 
consider some \cite{trichun,triothers} or all \cite{triour} of the 
dimensionless lepton number violating
couplings ($\lambda, ~\lambda'$ ) to be present in the superpotential. 
Several features of the neutrino mass spectrum like hierarchy and large
mixing are still preserved within these models making them phenomenologically
viable. However, unlike the bilinear model considered in this work, these
models are less constrained simply due to the large number of additional 
couplings present in the model. This can lead to the large solar neutrino
mixing angle even without relaxing the universality constraints on the
soft spectrum \cite{trichun,triour}. 

\section{Appendix A}

The renormalization group equations for various parameters
appearing in the soft scalar potential are basis dependent. We
have chosen a specific basis in which bilinear terms in the
potential are  kept in the superpotential till the weak scale.
These terms are rotated only after evolving to weak scale. We
collect here RG equations for relevant parameters with this
specific choice. They differ for example from the ones derived in
\cite{chun1loop} where relevant rotation is performed at each scale.
The following equations follow in a straightforward manner from
the formalism given by Falck\cite{falck}:
\bea
\label{mnuh1rge}
{d \over dt}m_{\nu_i'H_1'}^2 &=&m_{\nu_i'H_1'}^2
(- {1 \over 2} Y_i^{E}  - {1 \over 2} Y_\tau  - {3 \over 2} Y_b) ~, \\
{d\over dt}(\Delta m_i^2)&=& 3 Y_b ( m^2_{Q_3} + m^2_{D_3} + m^2_{H'_1}
+ \tilde{A}_b^2 )  \nonumber \\
\label{dmlrge}
&-& Y^E_i ( m_{L_i}^2 + m_{E_i}^2 + m_{H'_1}^2 + \tilde{A}^{E~2}_i ), \\
\label{dblrge} {d\over dt}(\Delta B_i)&=& \tilde{A}_\tau Y_\tau +
3 Y_b \tilde{A}_b - 3 Y^E_i \tilde{A}^E_i . \eea In the above, we
have used standard notation for all the soft parameters appearing
in the equations with the exception of $\tilde{A}$ which represents
the soft trilinear couplings.  
\section{Appendix B}
In this appendix we justify  the neglect of additional
contributions to neutrino masses not included in the main text. We
also discuss flavour violating processes $\mu\rightarrow e \gamma$
and show that the corresponding branching ratio is very small in
the present context.

Detailed analysis of the additional 1-loop diagrams contributing
to  neutrino mass matrix has  been done in
\cite{hempfling,romao,chun1loop}. While Refs.
\cite{hempfling,romao} calculate all the 1-loop self-energy
diagrams to the $7~ \times~ 7$ neutrino-neutralino mass matrix and
re-diagonalise it, Ref.\cite{chun1loop} follows the effective
mixing matrix approach. In addition to the contributions
considered in the text, large contributions are also expected from
diagrams which are not Yukawa suppressed, thus involving only
gauge vertices. These can be visualized as diagrams with two
R-parity violating mass insertions proportional to $\Delta
m_i^2,~\Delta B_i$ as given in eq.(\ref{scalar}), with neutralino
(chargino), sneutrino (charged slepton) and neutral Higgs (charged
Higgs) in the loops \cite{kaplan,chun1loop}. Typical magnitude
of these diagrams is given by

\be  \label{gsqdiag} {\cal M}^\l_{ij}\approx {g^2\over 16 \pi^2}
\e_i \e_j
 k'_ik'_j m_{susy}^{-1}\ee with
  $$k'_i\approx{c_1\Delta m_i^2+c_2\Delta B_i^2\over m_{\snui}^2}~.$$

 $m_{susy}$ is a typical supersymmetry breaking scale and $c_{1,2}$
 are coefficients of order one following from the scalar mass matrices
 of the model. $k'_i$ are similar to parameters $k_i$ defined in eq.(\ref{ki}).
 It is natural then to choose $k_i'\sim{\m\over v_1}k_i$ for order of magnitude estimates.
 Comparing the 1-loop gaugino contribution with the $b$-quark contribution
 (eq.(\ref{loop}))
 ${\cal M}^b$ we obtain

\be {{\cal M}^\lambda_{ij} \over {\cal M}^b_{ij}}\approx {g^2\over
16 \pi^2 A_b}\left({\mu\over v_1}\right)^2{k_ik_j\over  m_{susy}}
\ee

The numerical value of $A_b$ is given in eq.(\ref{abtau}). As
argued above, we typically need $k_{i}\sim 10^{-3} \doteq
10^{-4}$. It is seen that the $b - quark$ contribution retained in
the main text dominates over the gaugino contribution in this case
and it is consistent to neglect the latter. The other
contributions to neutrino masses are even less dominant than the
gaugino contribution\footnote{For a detailed discussion of the
various diagrams in mass insertion approximation, see
\cite{losada}.}. They  come from 1-loop diagrams with two Yukawa
vertices. These can be seen as a) diagrams with $\l$ and $h_\tau$
vertices with a R-parity violating mass insertion in the internal
line connecting charged slepton and charged Higgs, and b) diagrams
with $h_\tau$ couplings at both the vertices with two R-parity
violating mass insertions proportional to the sneutrino {\it vev}.
Both these sets of diagrams are suppressed by the $\tau$-Yukawa
coupling. They have been analyzed in detail in Ref.\cite{chun1loop}
where it has been shown that they can become comparable in
magnitude to $A_\tau$ in large tan $\beta$ regions. However as we
have seen earlier this contribution is always sub-dominant
compared to the contribution  from bottom Yukawa couplings, $A_b$.
Thus it is justified to neglect these contributions within the
present analysis.

\noindent {\bf Effects of Basis Rotation up to higher order in
$\e$ }: We now generalize the basis (\ref{basis1}) to higher order
in $\e$ and discuss its consequences. Such generalization becomes
necessary for discussion of  flavour violating transitions such as
$\m\rightarrow e \g$.  Eq.(\ref{basis1}) can be re-rewritten as
follows:
\be  \label{basis2} \bmat{c} H_1\\L_1\\L_2\\L_3\\ \emat
=\bmat{cccc} 1- \frac{1}{2}~ \ehat^2&\ehat_1&\ehat_2&\ehat_3\\
-\ehat_1&1-\frac{1}{2}~ \ehat_1^2&-\frac{1}{2}~ \ehat_1\ehat_2&-\frac{1}{2}~\ehat_1\ehat_3\\
-\ehat_2&-\frac{1}{2}~ \ehat_1\ehat_2&1-\frac{1}{2}~\ehat_2^2&-\frac{1}{2}~\ehat_2\ehat_3\\
-\ehat_3&-\frac{1}{2}~\ehat_1\ehat_3&-\frac{1}{2}~\ehat_2\ehat_3&1-\frac{1}{2}~\ehat_3^2\\
\emat ~ \bmat{c} H'_1\\L'_1\\L'_2\\L'_3\\ \emat,+\ord(\ehat^3) \ee
 where
$\ehat_i=(\e_i)/\m,\ehat^2=\ehat_1^2+\ehat_2^2+\ehat_3^2$. The
$V_{soft}$ in eq.(\ref{soft}) assumes the form
\bea \label{vsoft2} V_{soft}&=& m_{H_1}^2 |H_1^0 |^2 + m_{H_2}^2
|H_2^0|^2 + m_{\snui}^2 |\snui|^2  + \Delta m_i^2 \ehat_i
\left( \snui^{\star} H_1^0 + c.c \right)+(-\m B H_1^0 H_2^0+c.c) \nonumber \\
 &-&  \e_i \Delta B_i
\left( \snui H_2^0 + c. c \right)
-\frac{1}{2}\sum_{i<j}\ehat_i\ehat_j(\Delta m_i^2+\Delta
m_j^2)(\snui^*\tilde{\nu}_j+c.c.) ~, \eea
where
\bea m_{H_1}^2=m_{H_1'}^2 (1-\ehat^2)+m_{\snui'}^2 \ehat_i^2 ,
\nonumber \\
m_{\snui}^2=m_{H_1'}^2 \ehat_i^2+m_{\snui'}^2(1- \ehat_i^2),
\nonumber \\
B=B_\m (1-\ehat^2)+B_i \ehat_i^2 .\eea The rotation has generated
off-diagonal flavour violating sneutrino mixing terms at
$\ord(\e^2)$. Since these terms conserve lepton number, they do
not directly contribute to the neutrino masses but lead to flavour
violating transitions such as $\m\rightarrow e \gamma$.

The rotation in eq.(\ref{basis2}) induces mixing among the charged
leptons which were diagonal to start with. Define the charged
lepton mass matrix as

$$\L_i~M_l~e^c~,$$
then \be \label{charl}
M_l=\bmat{ccc} h_1d_1&\ehat_1\ehat_2 h_2f_2&\ehat_1\ehat_3 h_3
f_3\\
\ehat_1\ehat_2 h_1f_1&h_2 d_2& \ehat_1\ehat_3 h_3 f_3\\
\ehat_1\ehat_3 h_1 f_1&\ehat_2\ehat_3 h_2 f_2&h_3 d_3\\
\emat ~,\ee where
$$ d_i\equiv v_1(1+\frac{1}{2} (\ehat_i^2-|\ehat|^2)+\ehat_i^2 k_i-\ehat_l^2
k_l~,$$
$$f_i\equiv \frac{1}{2}v_1+k_i~.$$
The $k_i$ appearing in above are defined in eq.(\ref{ki}) and they
signify sneutrino \textit{vev} contribution to the charged lepton
mass matrix. As argued in the text, $k_i$ are required to be small
$\sim (10^{-3}-10^{-4})$ in order to account for the correct
neutrino masses. It then follows that sneutrino \textit{vev}
contribution to each element in $M_l$ is suppressed compared to
the corresponding contribution of $v_1$. Thus this contribution
can be neglected while diagonalizing $M_l$ in any realistic
theory. Even after neglecting it, the $\ord(\ehat^2)$ contribution
does produce additional mixing among charged leptons that is not
{\it Yukawa suppressed}. This is easily seen in the simplified
case of two generation. The $2\times 2$ version of the charged
lepton mass matrix is obtained from eq.(\ref{charl}) by setting
$\e_3=0$. The following rotation on the basis $(e_1,e_2)$ is
needed to diagonalize the charged letpon masses:
\be \label{emu} \bmat{c}e\\\m\\ \emat=\bmat{cc} 1&\frac{1}{2}
\ehat_1\ehat_2\\-\frac{1}{2}\ehat_1\ehat_2&1 \emat ~\bmat{c}e_1\\ e_2\\
\emat~.\ee $(e,\mu)$ here refers to the flavour basis. This
additional rotation affects the neutrino mixing terms in
eq.(\ref{vsoft2}) which can be re-written in the flavour basis as
\bea V_{soft}&=& m_{H_1}^2 |H_1^0 |^2 + m_{H_2}^2 |H_2^0|^2 +
m_{\tilde{\nu}_1}^2 |\tilde{\nu}_e|^2+m_{\tilde{\nu}_2}^2
|\tilde{\nu}_\m|^2 \nonumber
\\
&+& \left( \Delta m_1^2 \ehat_1 \tilde{\nu}_e^{\star} H_1^0 +
\Delta m_2^2 \ehat_2
\tilde{\nu}_\m^{\star} H_1^0+c.c \right) \nonumber \\
&-& \left( \m B H_1^0 H_2^0 + c. c \right) - \left( \e_1 \Delta
B_1 \tilde{\nu}_e H_2^0 +\e_2 \Delta B_2
\tilde{\nu}_\m H_2^0 + c. c \right) \nonumber \\
&-&\ehat_1\ehat_2 \frac{1}{2}(\Delta m_1^2+\Delta
m_2^2-m_{\tilde{\nu}_2}^2+m_{\tilde{\nu}_1}^2)(\tilde{\nu}_e^*\tilde{\nu}_\m+c.c.)
\label{vsoft3} .\eea

One sees that there are no additional lepton number violating mass
terms other than present at $\ord (\ehat)$. Thus discussion on
additional contribution to neutrino masses just given remains
unchanged. However, eq.(\ref{vsoft3}) contains lepton conserving
but flavour violating contribution proportional to
$\tilde{\nu}_e^*\tilde{\nu}_\m$. This can lead to process such as
$\m\rightarrow e\gamma$. The branching ratio for this process is
given by
\be BR(\m\rightarrow e\gamma)={12 \pi^2\over G_F^2 m_\mu^2}
|B|^2~. \ee
In the present case, the amplitude $B$ arises due to insertion of
the flavour violating sneutrino mass term given in the last term
in eq.(\ref{vsoft3}). This is approximately  given by
\cite{carlosemu}
\be |B| \sim {e^3 \over 16 \pi^2} {m_\mu\over
m_{\tilde{\nu}}^2 } \frac{1}{2} {\e_1\e_2\over \mu^2} k ~,\ee
where $k$ is a typical magnitude of $k_i$ and $m_{\tilde{\nu}}^2$
is sneutrino (mass)$^2$. As already argued, we need $\e_i\sim 0.1
~\GeV$ and $k\sim 10^{-3}$. Given this, last equation is seen to
give very small contribution to $ BR(\m\rightarrow e\gamma)~ \sim
{\cal O}(10^{-13}~|\e|^4 |k|^2 )$ which makes it unobservable in both
present \cite{mega} and future \cite{psi} experiments.

\noindent{\bf Acknowledgements:} We gratefully acknowledge useful
discussions with E. J. Chun which helped in clarifying several
issues connected with this work.


\newpage
\begin{figure}[h]
\centerline{\psfig{figure=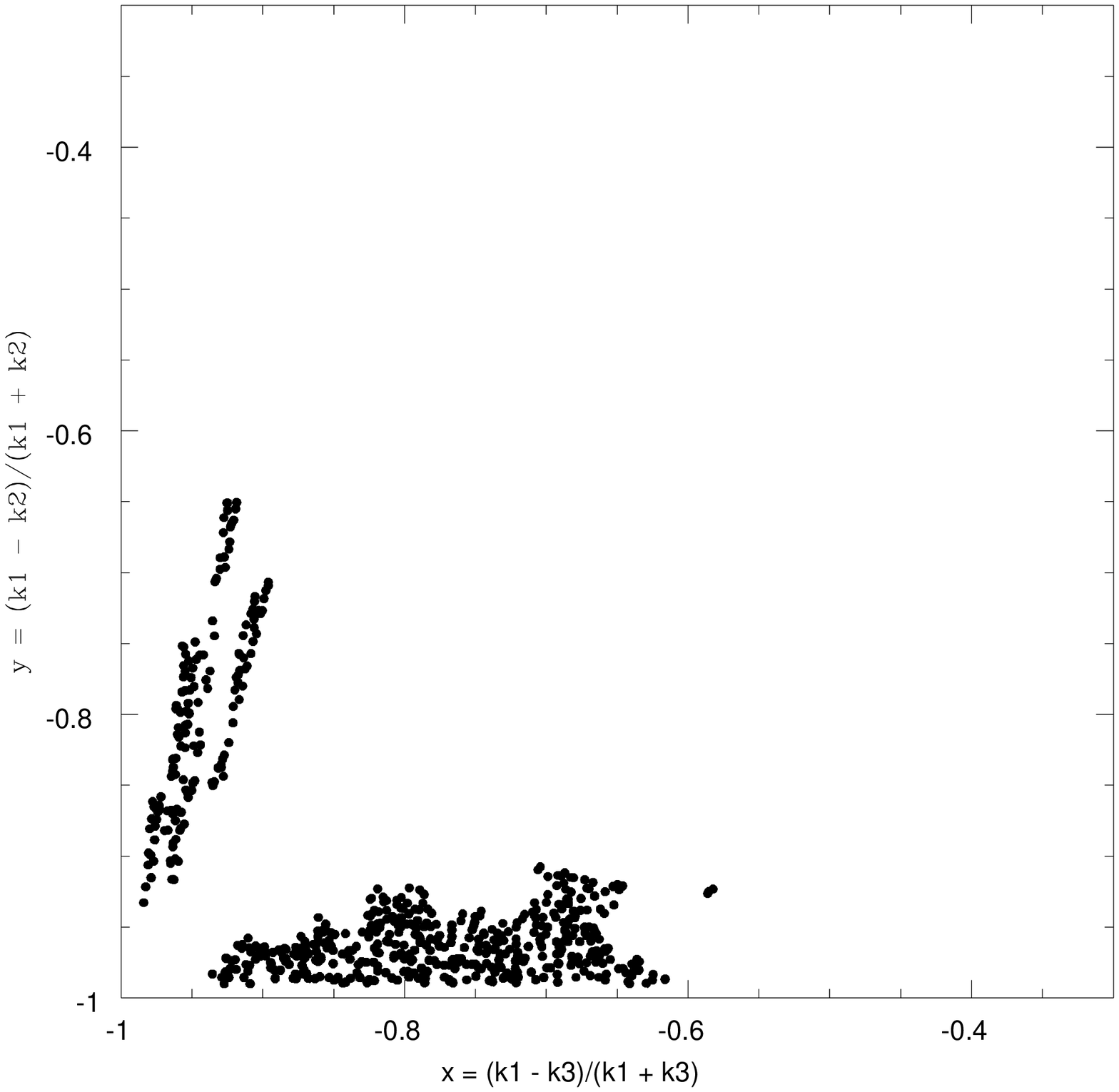,height=10cm,width=15cm}}
\caption{Allowed values of $x$ and $y$ for which all the neutrino
oscillation constraints are satisfied. The input values of
parameters are chosen in a way described in the text. }
\end{figure}

\end{document}